\documentclass[runningheads]{llncs}
\usepackage{graphicx}
\usepackage{diagbox}
\usepackage[table]{xcolor}
\usepackage{subcaption}
\usepackage{cite}
\usepackage{amssymb}
\usepackage{slashbox}
\usepackage{multirow}
\usepackage{adjustbox}
\begin{document}
\title{FedDis: Disentangled Federated Learning for Unsupervised Brain Pathology Segmentation\vspace{-5mm}}
\author{Cosmin I. Bercea\inst{1} \and
Benedikt Wiestler\inst{4} \and
Daniel Rückert\inst{2,3,4} \and
Shadi Albarqouni\inst{1,2}}
% %
\authorrunning{C. I. Bercea et al.}
\institute{Helmholtz AI, Helmholtz Center Munich, Neuherberg, Germany \and
Technical University of Munich (TUM), Munich, Germany \and
Imperial College Londen, UK \and
Klinikum Rechts der Isar, Munich, Germany}
\maketitle              % typeset the header of the contribution
\vspace{-7mm}
%
%
%
%%%%%%% ABSTRACT
%
\begin{abstract}
In recent years, data-driven machine learning (ML) methods have revolutionized the computer vision community by providing novel efficient solutions to many unsolved (medical) image analysis problems. However, due to the increasing privacy concerns and data fragmentation on many different sites, existing medical data are not fully utilized, thus limiting the potential of ML. Federated learning (FL) enables multiple parties to collaboratively train a ML model without exchanging local data. However, data heterogeneity (non-IID) among the distributed clients is yet a challenge.
To this end, we propose a novel federated method, denoted \textit{Federated Disentanglement} (\verb!FedDis!), to disentangle the parameter space into shape and appearance, and only share the shape parameter with the clients. \verb!FedDis! is based on the assumption that the anatomical structure in brain MRI images is similar across multiple institutions, and sharing the shape knowledge would be beneficial in anomaly detection.
In this paper, we leverage healthy brain scans of 623 subjects from multiple sites with real data (OASIS, ADNI) in a privacy-preserving fashion to learn a model of normal anatomy, that allows to segment abnormal structures.
We demonstrate a superior performance of \verb!FedDis! on real pathological databases containing 109 subjects; two publicly available MS Lesions (MSLUB, MSISBI), and an in-house database with MS and Glioblastoma (MSI and GBI).
\verb!FedDis! achieved an average dice performance of 0.38, outperforming the state-of-the-art (SOTA) auto-encoder by $42\%$ and the SOTA federated method by $11\%$. Further, we illustrate that \verb!FedDis! learns a shape embedding that is orthogonal to the appearance and consistent under different intensity augmentations.
\end{abstract}
%
%
%
%
%%%%% INTRO
%
%
\section{Introduction}
\begin{figure}[t]
\includegraphics[width=\textwidth]{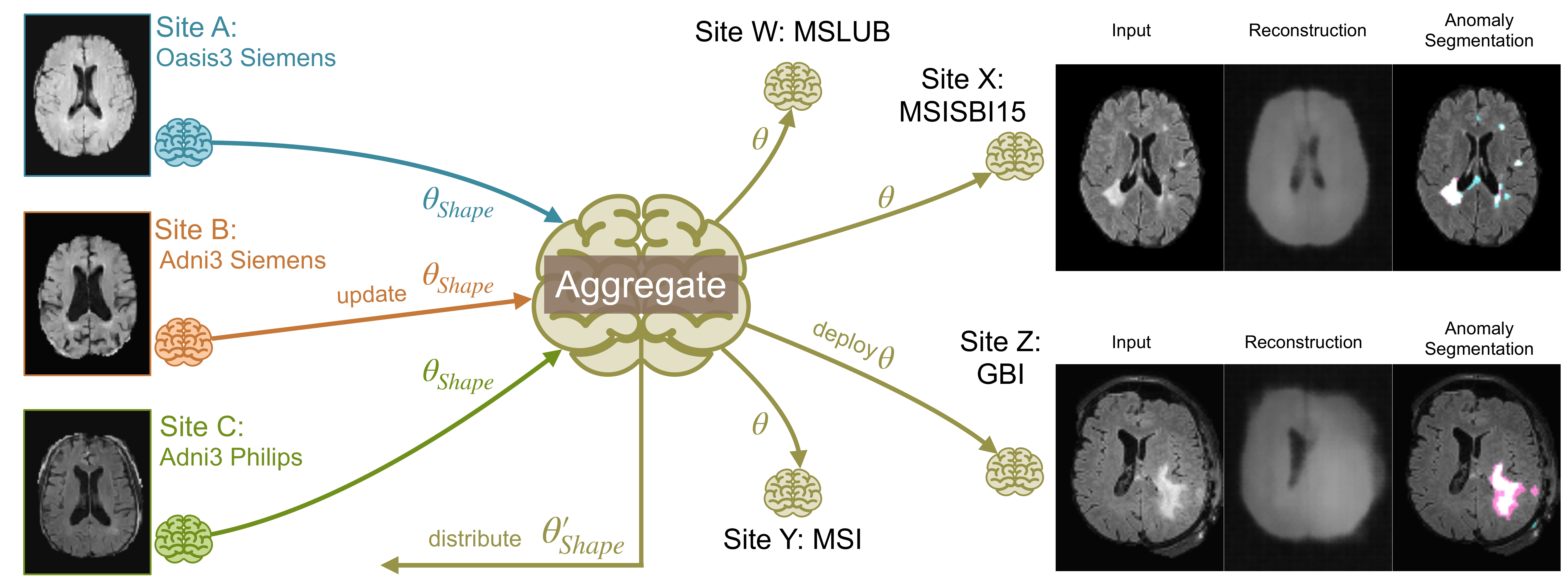}
\caption{Our FedDis framework: we model the healthy anatomy by leveraging brain scans from sites A to C in a privacy-preserving fashion. FedDis disentangles the model parameters into shape and appearance, and only share the shape parameters mitigating the data heterogeneity among the clients. The global FL model is then employed to segment anomalies on unseen sites W to Z.}\label{fig::1}
\end{figure}
Machine learning methods have shown promising results in various medical applications~\cite{litjens2017survey}. However, their performance highly depends on the amount and diversity of training data~\cite{sun2017revisiting}. In the context of medical imaging, is often infeasible to collect patient data in a centralized data lake due to privacy regulations~\cite{rieke2020future,kaissis2020secure}. Furthermore, medical data-sets are often siloed across many institutes and highly unbalanced due to the low incidence of pathologies or biased towards local demographics.
Federated learning (FL)~\cite{mcmahan2017communication} is a decentralized and privacy-preserving machine learning technique that showed promising results in medical imaging~\cite{albarqouni2020domain,li2019privacy,li2020federated,li2020multi,sarma2021federated,sheller2018multi, yang2021federated}.
Despite its advantages with regards to data privacy, the statistical heterogeneity in the data present at the different distributed clients negatively affects the federated performance~\cite{Konecny2016federated, li2020multi, mcmahan2017communication, yang2021federated}. This is especially the case for medical data which is inherently heterogeneous due to e.g., acquisition parameters, different manufactures of medical devices, and local demographics. Recent methods were proposed to tackle data heterogeneity and domain shifts (non-IID): FedVC~\cite{hsu2020federated} limits the number of local iterations performed at each client to avoid overfitting. On a different note, SiloBN~\cite{andreux2020siloed} and FedGN~\cite{hsieh2020non} improve the performance of federated methods in non-IID scenarios by not averaging local BN statistics or by using group normalization, respectively. Based on a similar observation, Li et al.~\cite{li2019privacy} re-initialize the momentum of ADAM optimizer at every local training round. In this work, we propose to tackle domain shifts based on the assumption that the anatomical structure in brain MRI images is similar, whereas the intensity distribution is somehow different across multiple institutions.
Inspired by recent works on disentangled representations~\cite{chartsias2019disentangled, locatello2019assumptions, sarhan2020fairness, shu2018deforming, qin2019unsupervised}, we propose a novel federated method, denoted \textit{Federated Disentanglement} (\verb!FedDis!) and argue its suitability for medical imaging. \verb!FedDis! disentangles the parameter space to shape and appearance, and only shares the shape parameters with the other distributed clients to train a domain-agnostic global model (\emph{cf.} Fig.~\ref{fig::1}).
Our main \textbf{contributions} are:
\begin{itemize}
    \item We introduce a \textbf{novel framework} of federated unsupervised anomaly segmentation by leveraging brain scans from multiple sites and databases using \textit{privacy-preserving federated learning.}
    \item We propose a \textbf{novel method, \textit{Federated Disentanglement}} (\verb!FedDis!) to disentangle the model parameters into shape and appearance and only share the shape parameters to mitigate the domain shifts of individual clients.
    \item We experimentally validate on multiple clients with real pathological databases showing a superior performance of \verb!FedDis! over the state-of-the-art. Specifically, \verb!FedDis! \textbf{significantly outperforms} FedVC~\cite{hsu2020federated}, SiloBN~\cite{andreux2020siloed} and FedGN~\cite{hsieh2020non} in addition to the best local- and data-centralised models~\cite{baur2021autoencoders}.
    \item \textbf{ We illustrate the disentanglement property} of our proposed  method. In particular we show that \verb!FedDis! learns a shape embedding orthogonal to appearance and consistent under different intensity shifts.
\end{itemize}
%%
%
%
%
%%%%%% METHODOLOGY
%
%
\section{Methodology}
The main concept behind our federated unsupervised anomaly segmentation framework, depicted in Fig.~\ref{fig::1}, is to model the distribution of healthy anatomy by learning how to efficiently compress and encode healthy brain scans from multiple institutions, then learn how to reconstruct the data back as close to the original input as possible. This enables the detection of pathology from faulty reconstructions of anomalous samples. We first formally introduce the problem and define the federated unsupervised anomaly segmentation setup. Then, we present \verb!FedDis! and elaborate a loss to enforce the disentanglement.
\subsection{Problem Formulation}
Given $M$ clients $\mathcal{C}_{j}$ with local database $\mathcal{D}_{j} \in \mathbb{R}^{H\times W\times N_{j}}$ consisting of $N_{j}$ healthy brain scans $\textit{\textbf{x}}\in \mathbb{R}^{H\times W}$, our objective  is  to  train  a  global  model $f(\cdot)$, in a privacy-preserved fashion, leveraging the healthy brain scans from multiple institutes, to segment the pathology \textit{$\textbf{r}_q$} $\in \mathbb{R}^{H\times W}$ for a given query brain scan \textit{$\textbf{x}_q$}.
\subsection{Federated Unsupervised Anomaly Segmentation}
FL aims to collaboratively learn a global model $f_{\theta^{G}}(\cdot)$, in a privacy-preserved fashion, without centralizing training data. Let $G$ be a global model and $\mathcal{C} = \{\mathcal{C}_{j}\}_{j=1}^M$ be a set of clients. At each communication round, the local clients are initialized with the global weights and trained locally on their own data sets for a fixed number of epochs to minimize following objective function:
\begin{equation}
    \min_{\theta^{G}}\mathcal{L}(\mathcal{D}; \theta^{G}) \quad \textrm{with} \quad  \mathcal{L}(\mathcal{D}; \theta^{G}) =\sum_{j=1}^{M}w_j\;\mathcal{L}_{Rec}(\mathcal{D}_{j}; \theta^{C_j}),
    \label{eq:formalism}
\end{equation}
where the learned local parameters $\theta^{C_j}$ are aggregated to a new global model: $\theta^G \leftarrow \sum_{j=1}^M w_j\; \theta^{C_j}$, where $w_j=\frac{N_j}{\sum_{j=1}^{M} N_j}$ is the respective weight coefficient.
A popular architecture to learn efficient data encoding in an unsupervised manner are convolutional auto-encoders (AE)\cite{vincent2010stacked,baur2021autoencoders} where  an encoder is trained to compress $\textit{\textbf{x}}$ to a latent representation $\textit{\textbf{z}} \in \mathbb{R}^d$, from which a decoder attempts to reconstruct the original by minimizing following objective:
\begin{equation}
    \arg\min_{\theta^{C_j}}\sum_x \mathcal{L}_{Rec}(x, x_{Rec}), \hspace{5pt}\textrm{with}\hspace{5pt} x_{Rec}=f_{\theta^{C_j}}(x),
\end{equation}
where a common choice for the reconstruction loss is the mean absolute error: $\mathcal{L}_{Rec}(x,x_{Rec}) = \frac{1}{N} \sum_{i=0}^N | x-x_{Rec}|$.
Finally, the anomaly segmentation is given by the binarization of the residual $r = x - x_{Rec}$. Yet, the performance is negatively affected by data heterogeneity and domain shifts among the clients.
\subsection{Federated Disentanglement (FedDis)}
% Figure
\begin{figure}[tb]
     \centering
     \begin{subfigure}[b]{0.495\textwidth}
         \centering
         \includegraphics[width=\textwidth]{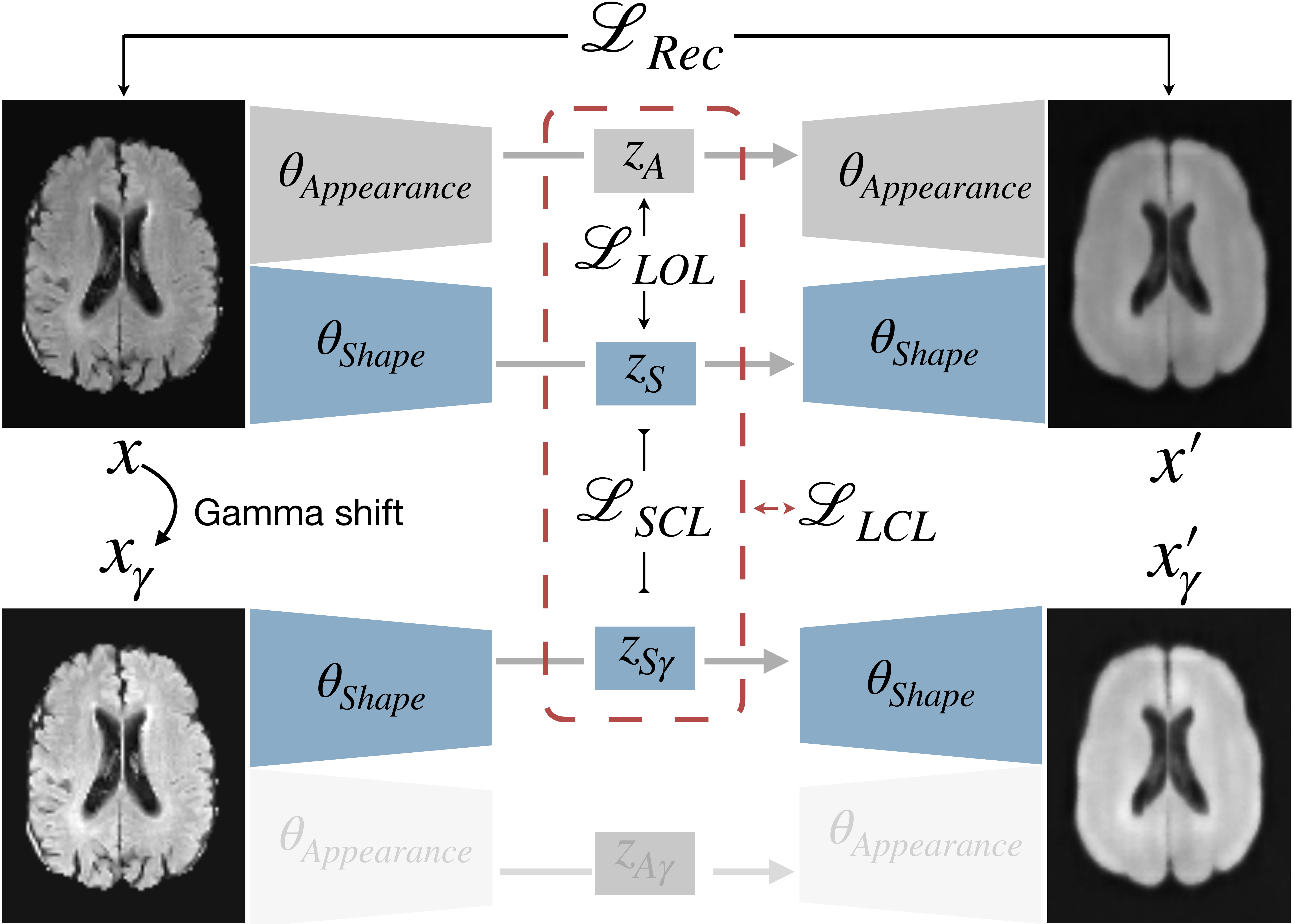}
         \caption{Disentangled Auto-encoder}
         \label{fig::2a}
     \end{subfigure}
     \hfill
     \begin{subfigure}[b]{0.495\textwidth}
         \centering
         \includegraphics[width=\textwidth]{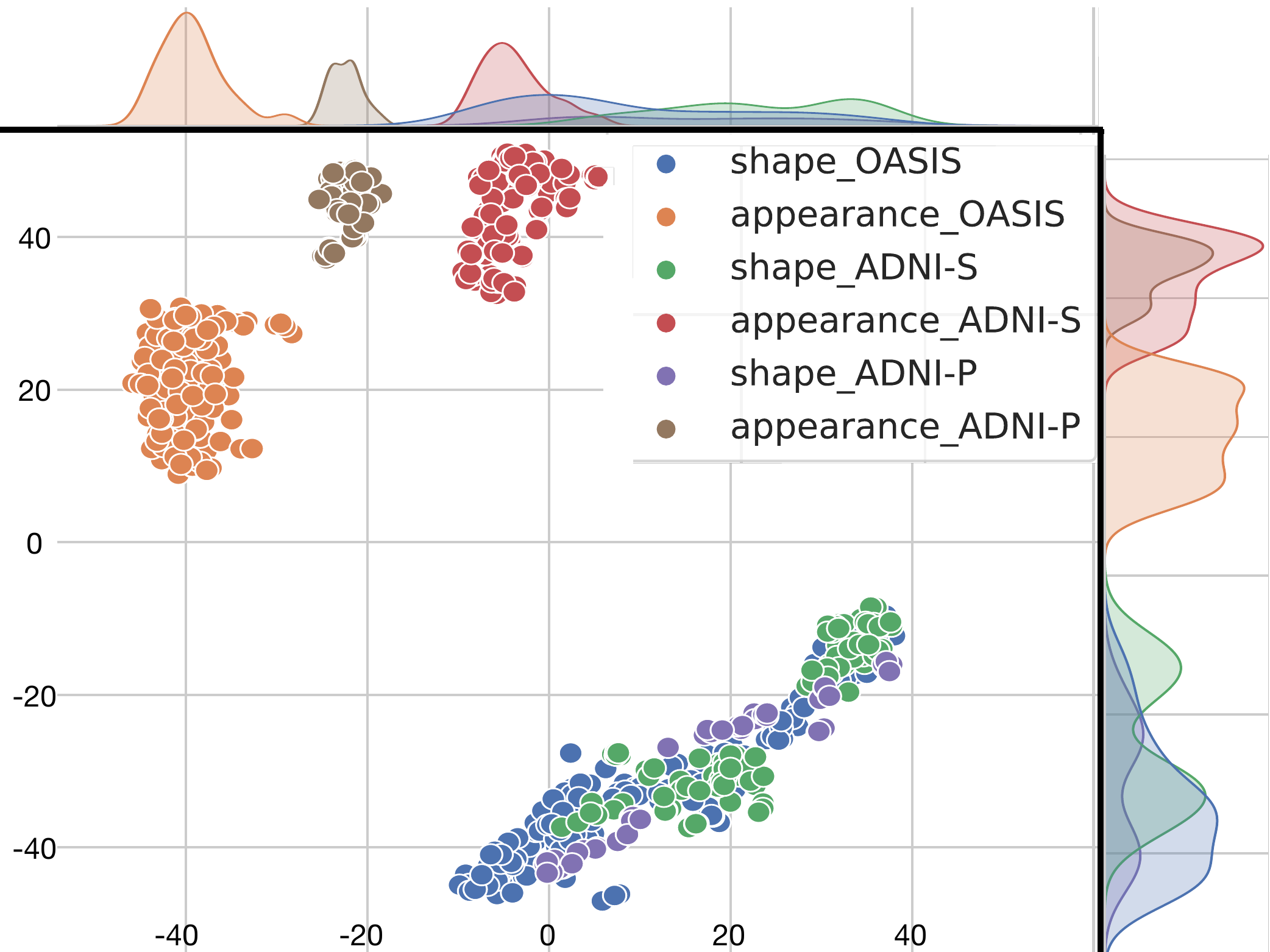}
         \caption{Latent embedding with t-SNE\cite{van2008visualizing}}
         \label{fig::2b}
     \end{subfigure}
     \caption{(a) shows our disentangled architecture and the latent loss terms. In (b) FedDis learns a latent embedding, where the appearance of the clients is well seperated, while the shape parameters follow the same distribution.}
     \label{fig::2}
\end{figure}
We make following assumptions: (i) the statistical heterogeneity among the distributed clients is mostly due to different scanners and different acquisition parameters and (ii) the anatomical structure information is alike across different institutes and acquisition parameters. In this light, we propose to disentangle the model parameters $\theta$ into shape $\theta_{S}$ and appearance $\theta_{A}$ as seen in Figure~\ref{fig::2a}. This allows us to leverage the shared mutual structural information from different clients and mitigate the domain shifts by allowing personalized appearance models. After every communication round, the global model is averaged as follows: $\theta^G \leftarrow \sum_{k=1}^K w \theta^{Lk}_{S}$. To enforce the disentanglement, we introduce following requirements: (i) shape consistency (SCL): shape embeddings should be similar under different intensity augmentations e.g., changes in brightness/ contrast or random gamma shifts - we choose the latter; and (ii) latent orthogonality (LOL): distributions of latent representations for shape and appearance should be different. Thus, we define the latent contrastive loss (LCL) as:
\begin{equation}
   \mathcal{L}_{LCL}(z_A,z_S,z_{\gamma S};\theta_S;\theta_A) = \beta \underbrace{\mathcal{L}_{KL}(z_S, {z_{\gamma S}};\theta_S)}_{\mathcal{L}_{SCL}} + (1-\beta) \underbrace{(1-\mathcal{L}_{KL}(z_A, z_S;\theta_A)}_{\mathcal{L}_{LOL}},
\end{equation}
where $z_A, z_S$ and $z_{\gamma S}$ are the latent representation for the appearance and shape and shape of the gamma shifted image, respectively; $\beta$ is used to weigh the two contrasting loss terms, and $\mathcal{L}_{KL}(z_A, z_S;\theta_A) = \textrm{KL}(p(z_A|x;\theta_A)|| p(z_S|x;\theta_S))$ is the KL divergence. The overall objective is given by:
$\alpha\mathcal{L}_{Rec}(x,  x_{Rec})  +  (1-\alpha)\mathcal{L}_{LCL}(z_A, z_S,z_{\gamma S})$, with weight $\alpha$.
\section{Experiments and Results}
We experimentally validated our method on multiple clients with real pathological databases. Our main findings are i) The domain shifts among the clients negatively affects the performance of both data-centralized and FL methods. ii) \verb!FedDis! mitigates the domain shifts and significantly improves the performance and iii) \verb!FedDis! successfully disentangles the appearance and shape embeddings.
\subsection{Experimental Setup}
%
% TABLE
\begin{table}[t]
    \centering
   % \rowcolors{1}{}{gray!10}
    \setlength{\tabcolsep}{4pt}
	\caption{Databases. We split the data randomly, patient-wise into training, validation and testing for healthy subjects and segment anomalies on multiple sclerosis (MS) and glioblastoma (GB) sequences.}\label{tab::1}
%    \begin{adjustbox}{width=\textwidth,center}
\resizebox{\textwidth}{!}{%
    	\begin{tabular}{l |c c c c c c c}
    	    \hline
    	    & OASIS-3~\cite{LaMontagne2019oasis} & ADNI-S~\cite{Weiner2016adni} & ADNI-P~\cite{Weiner2016adni} & MSLUB~\cite{lesjak2018mslub} & MSISBI~\cite{CARASS2017msisbi} & MSI & GBI\\
    		\hline\hline
    		Train/Val/Test & 376/124/135 & 303/99/105 & 138/42/48 & -/-/30 & -/-/21 & -/-/48 & -/-/97\\
    		Cohort & Healthy & Healthy & Healthy & MS & MS & MS & GB\\
    		Scanner (3.0 Tesla) & Siemens & Siemens & Philips & Siemens& Philips & Philips & Philips\\
    		Image size/resolution& \multicolumn{7}{c}{$128\times128px$ / $2\times2mm$}\\
    		\hline
    	\end{tabular}
%    \end{adjustbox}
}
\end{table}
\paragraph{Datasets and implementation.}
We used two publicly available brain MR datasets (OASIS-3 and ADNI-3) for training and two publicly available MS lesion datasets (MSLUB, MSISBI) as well as an in-house database with MS and Glioblastoma (MSI, GBI) for testing, as seen in Table~\ref{tab::1}.
For the local, data-centralized and federated baseline experiments, we used an auto-encoder~\cite{baur2021autoencoders} with the number of filters from 32 to 128, a spatial bottleneck $\textit{\textbf{z}} \in \mathbb{R}^{8\times8\times128}$ and a dropout of 0.2. For \verb!FedDis!, we split the parameters in half for every path to not add more complexity to the network. We trained the model for 50 rounds, each with 5 local epochs with a batch size of 8. We used an ADAM~\cite{kingma2014adam} with a learning rate of $1^{-4}$ and exponential decay of 0.97. We set the loss weights $\alpha$ and $\beta$ to 0.2 and 0.5, respectively and the random gamma shift range to [0.5, 2].
\paragraph{Pre- and post-processing.}  All scans have been registered to the SRI24 atlas template space ~\cite{rohlfing2010sri24} to ensure all data share the same volume size and orientation. Subsequently, the scans have been skull-stripped with ROBEX~\cite{iglesias2011robex} and normalized to the [0,1] range. We used the axial mid-line with a size of 128$\times$128px and resolution of 2$\times$2mm. For post-processing, we follow \cite{baur2021autoencoders} by first multiply each residual image by a slightly eroded brain mask to remove noticeable residuals that occur near sharp edges at the borders of the brain. In addition, we use prior knowledge and keep only positive residuals, as these lesions are known to be fully hyper-intense in FLAIR images. Further, we apply a median filtering of size 3 to remove small outliers and obtain a more continuous signal. Finally, we binarize the results using their 99 percentile and perform connected component analysis to discard all small structures with an area less than 4 pixels.
\paragraph{Evaluation metrics.} We use the DICE score to measure the anomaly segmentation performance of the models. For the FL systems, we report the performance of the global model shared among the clients and for \verb!FedDis! we take the learned shape model and add the average appearance parameters (after training), in contrast to FedAvg where all the parameters are trained globally. We report the relative improvement (RI) of a over b as: $(a-b)/b$ and use the structure similarity SSIM to measure the reconstruction fidelity.
\subsection{Anomaly Segmentation}
% TABLE
\begin{table}[t]
    \centering
   % \rowcolors{1}{}{gray!10}
    \setlength{\tabcolsep}{2pt}
	\caption{Experimental results. Mean and std DICE results for anomaly segmentation. RI shows the relative improvement over~\cite{baur2021autoencoders} and the last two columns show mean SSIM results on the averaged healthy test sets and a healthy test set from an unseen distribution. */** mark statistical significant improvements over~\cite{baur2021autoencoders} and best federated method respectively (ks-test; p$\leq$0.05).}\label{tab::2}
    \begin{adjustbox}{max width=\textwidth,center}
    % \resizebox{\textwidth}{!}{%
    	\begin{tabular}{l| c c c c |c |c c}
    	    \hline
    	    \multirow{2}{*}{\backslashbox{Method}{Dataset}} &  MSLUB(Si) &  MSISBI(Ph) &  MSI(Ph)  & GBI(Ph) &  RI & Test & Healthy \\
    		%Scanner \& Sequence &  S3T-T1w & GE3T-T1w & P3T-T1w & S3T-Flair & GE1.5T-T1w & P3T-Flair & P3T-Flair & P3T-T1w\\
    		& \multicolumn{4}{c}{DICE$\uparrow$} & \% &\multicolumn{2}{c}{SSIM$\uparrow$} \\
    		\hline\hline
    		{Baur et al. (AE)~\cite{baur2021autoencoders}}: & & & & & & &\\
    		\enspace- In-house dataset
    		&0.098 $\pm$ 0.116 & 0.066 $\pm$ 0.073 & 0.165 $\pm$ 0.134 & 0.239 $\pm$ 0.127 & -45\% & N/A & N/A\\
    		\enspace- Local: OASIS(Si)
    		 &0.243 $\pm$ 0.179  & 0.154 $\pm$ 0.098 & 0.305 $\pm$ 0.168 & 0.262 $\pm$ 0.155 & -1\% & 0.849 & 0.809 \\
    		\enspace- Local: ADNI(Si)
    		&0.233  $\pm$ 0.182 & 0.124 $\pm$ 0.117 & 0.339 $\pm$ 0.188& 0.277 $\pm$ 0.180 &-3\% & 0.836 & 0,794\\
        	\enspace- Local: ADNI(Ph)
    		&0.238  $\pm$ 0.188 & 0.112 $\pm$ 0.108 & 0.264 $\pm$ 0.163& 0.255$\pm$ 0.169 & -13\% & 0.817 & 0.777\\
    		\enspace- Data-centralized
    		&0.234 $\pm$ 0.178  &0.150 $\pm$ 0.116 & 0.310 $\pm$ 0.173& 0.291 $\pm$ 0.170 & 0\% &\bfseries 0.860 &\bfseries 0.819\\
    		\hline
    		FedAvg~\cite{mcmahan2017communication}
    		&0.250 $\pm$ 0.191  & 0.190 $\pm$ 0.133 & 0.361 $\pm$ 0.194& 0.306 $\pm$ 0.165 & 14\% & 0.838 & 0.803 \\
    		FedVC~\cite{hsu2020federated}
    		&0.253 $\pm$ 0.204  & 0.160 $\pm$ 0.113 & 0.377 $\pm$ 0.190& 0.320 $\pm$ 0.179 & 12\% & 0.815 & 0.785 \\
    		SiloBN~\cite{andreux2020siloed}*
    		&0.269 $\pm$ 0.212  & 0.214 $\pm$ 0.173 & 0.372 $\pm$ 0.200& 0.358 $\pm$ 0.202 & 25\% & 0.838 & 0.803 \\
    		FedGN~\cite{hsieh2020non}*
    		&0.245 $\pm$ 0.197  & 0.195 $\pm$ 0.159 & 0.406 $\pm$ 0.223& 0.354 $\pm$ 0.181 & 22\% & 0.837 & 0.807 \\
    		\hline
    	    FedDis(ours)**
    		&\bfseries0.281 $\pm$ 0.234  &\bfseries 0.237 $\pm$ 0.220 & \bfseries0.424 $\pm$ 0.224&\bfseries 0.417 $\pm$ 0.203  & \bfseries 40\%  & 0.804 & 0.792 \\
    		\enspace- w/o LOL(ours)**
    		&\bfseries 0.272 $\pm$ 0.227  &\bfseries 0.242 $\pm$ 0.239 & \bfseries 0.420 $\pm$ 0.226 &\bfseries 0.411 $\pm$ 0.206 & \bfseries 39\% & 0.793 & 0.791 \\
    		\enspace- w/o SCL(ours)
    		&0.268 $\pm$ 0.204  & 0.190 $\pm$ 0.132 & 0.360 $\pm$ 0.191& 0.315 $\pm$ 0.166 & 16\% & 0.821 & 0.797 \\
    		\enspace- w/o LCL(ours)
    		&0.265 $\pm$ 0.210  & 0.204 $\pm$ 0.164 & 0.363 $\pm$ 0.191& 0.328 $\pm$ 0.164 & 20\% & 0.822 & 0.786 \\ \hline

    	\end{tabular}
    \end{adjustbox}%}
\end{table}
% FIGURE
\begin{figure}[t]
\includegraphics[width=\textwidth]{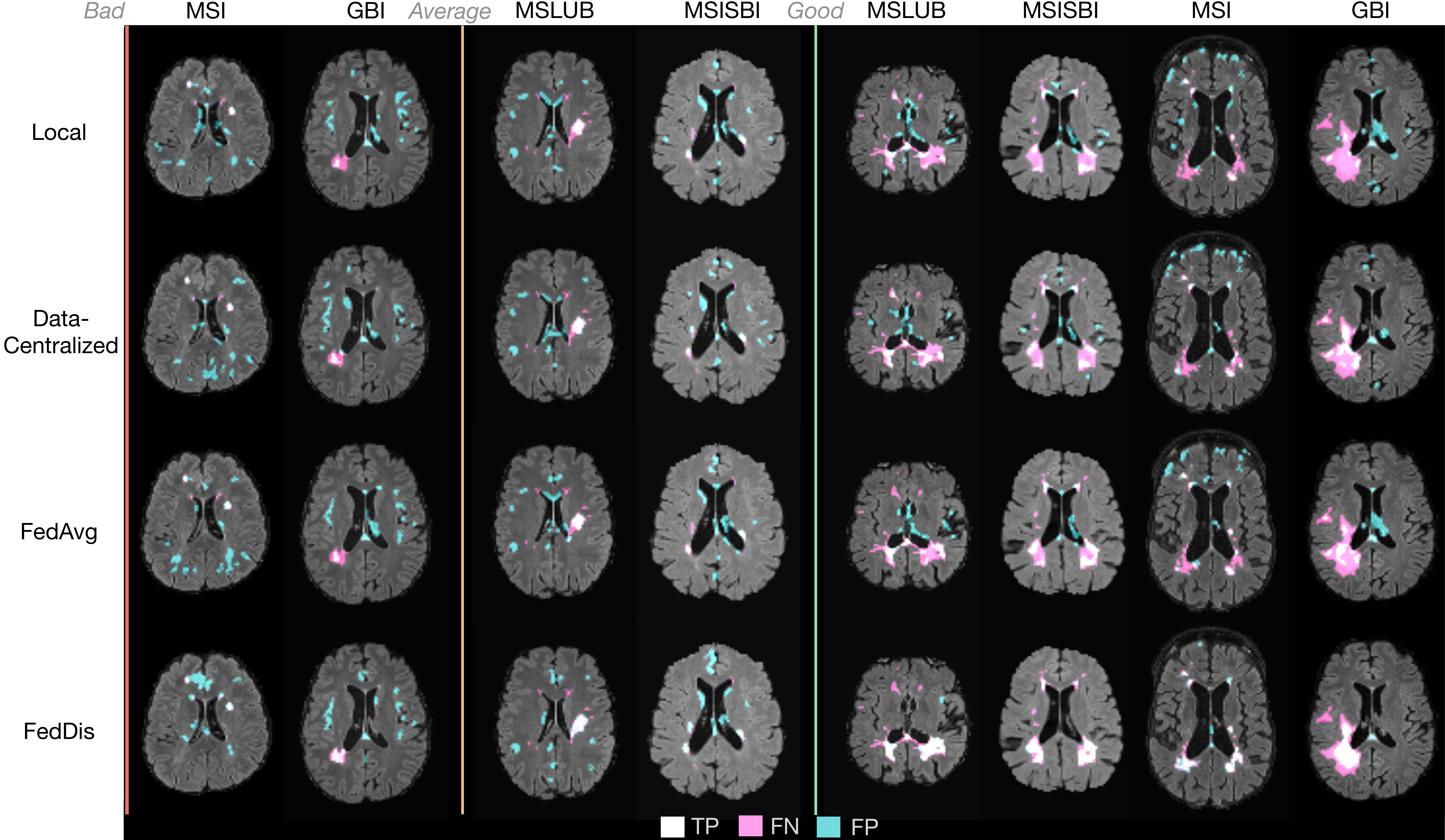}
\caption{Anomaly segmentation visualization. The rows show different methods and the columns show bad, average and good result from different data sets.}\label{fig::3}
\end{figure}
% Quantitative
Table~\ref{tab::2} summarizes quantitative results. We consider the following methods: Baur et al.~\cite{baur2021autoencoders}: auto-encoder trained on different local datasets; data-centralized: model trained with the local clients centralized in a data-lake; federated: FedAvg baseline and several federated methods that tackle data heterogeneity; and our proposed method, \verb!FedDis!. First, note that even though the data-centralized model did  achieve the best results in terms of reconstruction accuracy, it performed poorly on the anomaly segmentation task. Further, different variants of the FedAvg that tackle domain shifts are performing statistically significant better than Baur et al.~\cite{baur2021autoencoders}, suggesting there is a domain-shift among the distributed clients. Finally, \verb!FedDis! significantly outperforms Baur et al.~\cite{baur2021autoencoders} by 40\% and the best federated method, siloBN,  by 11\% on the main anomaly segmentation task with a marginal loss in reconstruction fidelity. The latent contrastive loss (LCL) plays hereby an essential role for the improved performance.
%
% Qualitative
Figure~\ref{fig::3} illustrates bad, average and good samples of anomaly segmentation results for different methods. \verb!FedDis! reduces the amount of false positive and negatives and has a more robust segmentation output for both MS and GB pathologies.
\subsection{Ablation: Domain shifts and disentanglement}
% Figure
\begin{figure}
     \centering
     \begin{subfigure}[b]{0.495\textwidth}
         \centering
         \includegraphics[width=\textwidth]{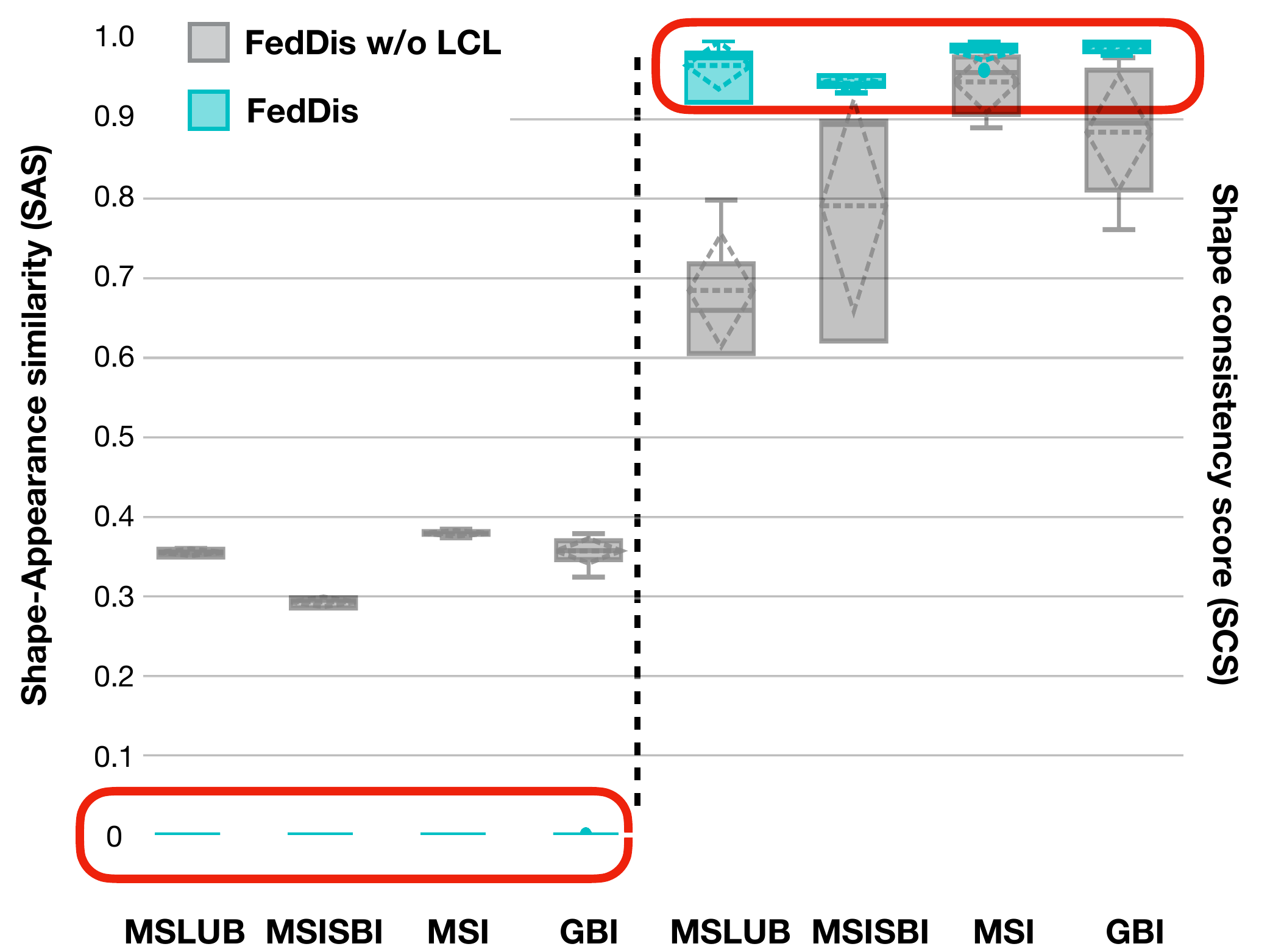}
         \caption{SAS$\downarrow$. SCS$\uparrow$. (cosine similarity)}
         \label{fig::4a}
     \end{subfigure}
     \hfill
     \begin{subfigure}[b]{0.495\textwidth}
         \centering
         \includegraphics[width=\textwidth]{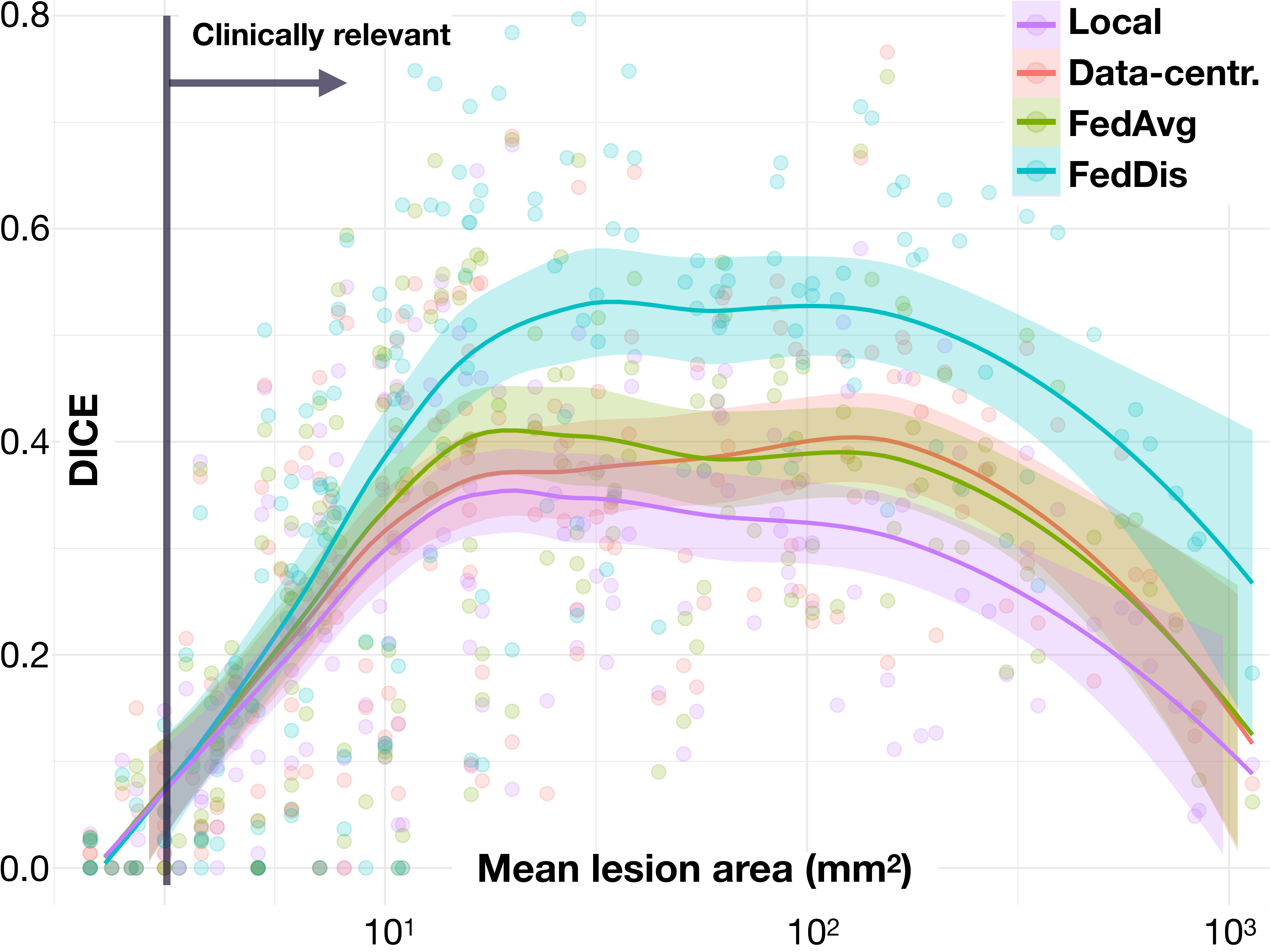}
         \caption{DICE vs. mean lesion area/slice.}
         \label{fig::4b}
     \end{subfigure}
     \caption{(a) shows that FedDis learns a latent shape representation that is orthogonal to the latent appearance representation and remains identical under different intensity changes. In (b) we show that FedDis outperforms the baseline for lesions that are clinically relevant ($>$ 3mm - current MS guidelines). }
     \label{fig::4}
\end{figure}
Here, we provide a detailed analysis of the disentanglement property of FedDis. Figure~\ref{fig::2b} shows an illustration of the latent representation of our method. The appearance of different clients is well separated, while the shape embeddings cluster together. This validates our assumption that the data heterogeneity in the clients is given by different appearances and that the shape parameters are similar across different institutes. The latent contrastive loss (LCL) is an essential part for the successful disentanglement. Figure~\ref{fig::4a} shows that FedDis, learns a shape representation that is orthogonal to the appearance and consistent under different intensity changes, in contrast to the naive disentanglement without LCL. However, the latent orthogonal loss helps only marginally and even worsens the results if used without the shape consistency loss, as seen in Table~\ref{tab::2}. This could be attributed to the unwanted effect that the appearance of different clients are indirectly enforced to be similar. Figure~\ref{fig::4b} shows the mean performance of FedDis compared to the baselines for different lesion sizes. Interestingly, FedAvg performs better than Baur et al.~\cite{baur2021autoencoders} for smaller lesions. FedDis significantly outperforms the other models for all clinically relevant lesions.
%
%
%%%
%
\section{Discussion and Future Work}
\paragraph{Generalizability to unseen scanners.}
Since most of the data sets with anomalies do not also provide healthy data, the question how to choose the appropriate trained appearance remains open. In our experiments, we used an averaged appearance model of the trained clients to perform anomaly segmentation on new/unseen data sets. This differs from the standard federated averaging where all model parameters are optimized globally during the training. As future work we want to investigate different aggregation methods for such cases. Furthermore, we plan to investigate the benefits of a client with anomalies contributing to the federated training with healthy data from the same distribution e.g., same scanner, thus enabling the use of a personalized appearance models for inference.
\paragraph{Reconstruction fidelity.} The data-centralized model achieved the best reconstructions, but performed rather poorly on segmentation pathologies. In contrast, despite the marginal drop in the reconstruction fidelity, different federated methods performed significantly better at segmenting pathologies. We hypothesize that considering the reconstruction error alone while training, ignores the models capacity to reconstruct pathologies, resulting in poor segmentations. In future work we plan to enforce the healthy reconstruction of difficult patches~\cite{bercea2019shamann}.
\section{Conclusion}
We proposed a new framework for federated unsupervised brain anomaly segmentation, which leverages scans from multiple sites and databases in a privacy-preserving manner. In this context, we introduced a new federated method, FedDis, to mitigate the domain shifts of distributed clients by disentangling the model parameters into shape and appearance and only share the shape with the other clients. We experimentally validated on real pathological datasets the superiority of \verb!FedDis! over SOTA AE and related federated methods.  In future work,we plan to add more clients to the federated setting and use different MRI sequences, e.g., T1-/T2-weighted sequences, for enforcing the shape consistency.
%
%
% %
%
% %
% % ---- Bibliography ----
% %
%\bibliographystyle{splncs04}
%\bibliography{main}

% %
%
\end{document}